# The Electronics Design of Error Field Feedback Control System in KTX

Tianbo Xu, Kezhu Song, Junfeng Yang

*Abstract*—KTX (Keda Tours eXperiment) is a new RFP (reversed field pinch) device at the University of Science and Technology of China. The unique double-C design of the KTX makes modifications and investigations of power and particle control easy, but the error field of slit zone in the new design should not be neglected. The objective of this paper is to introduce a new active feedback control system which can change the voltage between the unique double-C structures to make the toroidal field better. FPGA is the central part of the whole system to control all the process, because it can manipulate and transmit the data from coils in real time. There are 2 high-speed 8-channels ADCs in the system to convert the analog signal from 16 Rogowski coils which can detect dynamic eddy current of copper shells near the vertical gap. FPGA also control the external power amplifier to change the voltage between the unique double-C structures by commanding 16 high-speed DACs to give the RFP device a feedback. Result indicated that the error field in KTX device was reduced, and the system could successfully achieve fast matrix calculation with lower delay.

## I. Introduction

Reversed field pinch (RFP) device is a torus fusion device in which plasma is produced by external power supply system[1]. KTX (Keda Torus for eXperiment) is a medium size RFP (major radius R = 1.4 m, minor radius a = 0.4 m) [2,3]. The vacuum vessel and the conducting shell in KTX device are mounted on a movable platform, so that the two parts of the vacuum vessel can be moved in opposite directions. This structure, called 'double-C', allows the machine to be opened easily without dissembling the poloidal field windings around.

However, the gaps and flanges on the reversed field pinch device can split the stable copper shell, which will cut off and change the induced current path and form an error field. The error field can reduce the length of the magnetic field connection, weaken the plasma confinement performance. In addition, the error field may affect the quality of the plasma, reduce the duration of the discharge, and even directly cause "mode-lock". Due to Resonant Field Amplification(RFA), the error field can also influence plasma stability more significantly in some special situation. [4]According to the simulation, the error field of the vertical gap may lead to around one degree declination of signal-to-noise ratio, which cannot be tolerated. [5]

In this article, a feedback control system to solve the error field problem is presented. It is extremely important to build up a feedback system for this new fusion device. The main part of the system is two different boards with their independent function modules. Sample module is used to acquire and to manipulate the data from Rogowski coils around the gap. Through the FPGA on the sample module, the module can do some complicated matrix algorithm including PID control. After the calculation of data, the sample module transmits the result to the coil control module. Coil control module gets the data from sample module and gives a feedback signal to the active error field control power amplifier. FPGA is the core of both module, which can make a complicated matrix calculation and control the whole module.

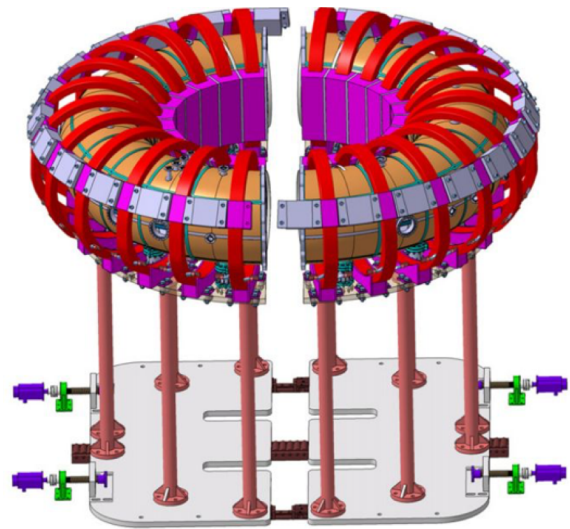

Fig.1. The 'double-C' structure

This work was supported University of Science and Technology of China

Tianbo. Xu is with State Key Laboratory of Particle Detection and Electronics, Department of Modern Physics, University of Science and Technology of China, Hefei, China, 230026, (e-mail:xtb305@mail.ustc.edu.cn).

Kezhu. Song is with State Key Laboratory of Particle Detection and Electronics, Department of Modern Physics, University of Science and Technology of China, Hefei, China, 230026, (e-mail: skz@ustc.edu.cn,).

Junfeng. Yang is with State Key Laboratory of Particle Detection and Electronics, Department of Modern Physics, University of Science and Technology of China, Hefei, China, 230026, (e-mail: yangjf@ustc.edu.cn,).

## II. Hardware Structure of System

### A. KTX fusion device

According to the structure of the KTX, Rogowski coils and power amplifiers have a long distance. A long-distance analog transmission can create a large amount of noise, so the whole system has to be divided into two parts to reduce the path of analog transmission. In the meanwhile, two different parts can make the process more unambiguous, which is necessary for the whole program.

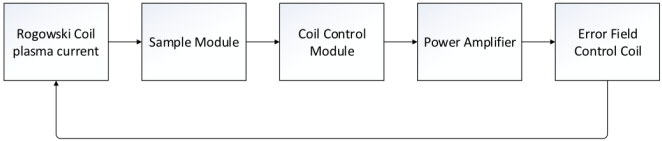

Fig.2. the Simplified Architecture of the feedback system

The system is composed of two different modules: the sample module and the coil control module. Sample. The main structure of the error field feedback system is shown in figure 2

### B. SAMPLE MODULE

The main task of the sample module is data sampling and data processing. Figure 3 shows the structure of the sample module. According to the parameter mentioned, ADS8528 is a proper choice to fulfill the target. The ADS8528 contains eight low-power 16-bit, successive approximation register (SAR) based analog-to-digital converters (ADCs) with true bipolar inputs. FPGA uses these high-speed ADCs to get the current signals which from 16 Rogowski coils. Due to the faintness of current signal, the analog signal will go through two-stage amplifiers to make it fulfill the measuring range of ADCs. Then FPGA would do some complicated matrix algorithm including PID control. The RS-485 is the protocol of transmission between two boards, which connects sample module with coil control module. RS-485 is intended to 40Mbps, and it is convenient to set up. There is also DDR2 and network interface in the sample module to store and verify the data from coils.

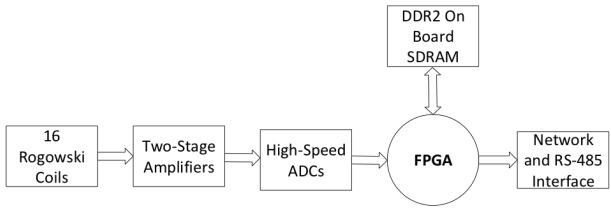

Fig.3. the Simplified Structure of Sample Module

### C. COIL CONTROL MODULE

In the coil control module, FPGA gets the processed data from RS-485 interface. Network Interface provides another option for receiving data and gives the data back to the computer to record initial data. DAC8831-EP is the DAC used in this module. It features 16-bit resolution and a standard high-speed (clock up to 50 MHz) SPI serial interface to communicate with FPGA, which assures high speed and proper accuracy of feedback signal. With the help of operational amplifier, DAC is able to generate analog signal whose range is from -5V to +5V.

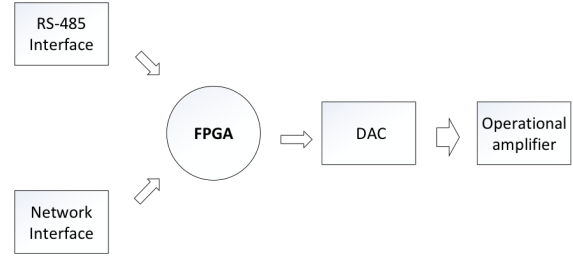

Fig.4. the Simplified Structure of Coil Control Module

## III. SYSTEM ARCHITECTURE

There are two essential algorithms in sample module. One is mutual inductance correction, the other is PID controller. It is obvious that the mutual inductance matrix only has three different parameters. So when FPGA gets a 12-bit data from one path, the data will go through two different multipliers which have signed coefficient. After sampling and calculating of all data from 16 paths in a period, FPGA will do the addition algorithm based on the matrix multiplication rules. Then the PID controller gets the data from mutual inductance correction. According to the discrete implementation of PID controller, it has their adjustable parameters and three multipliers. The adjustable parameters will be clarified through the fundamental test.

### A. Inductance correction

The new designed feedback control system is used in the KTX toroidal field which had some differences from other fusion device. There are 16 Rogowski coils around the gap from each side. And power amplifiers are installed in a power room. According to the simulation, the signal from Rogowski coils is a bipolar signal. The mutual inductance between two coils should not be ignored. The mutual inductance function is:

$$\overrightarrow{U_{OUT}} = v\vec{M} \cdot (\overrightarrow{U_{IN}} \cdot \beta_R + \alpha_0) \quad (1)$$

$U_{OUT}$, $U_{IN}$ are the voltage of the input and output. And $v$ is the reciprocal of amplification factor of power amplifier. $\alpha_0$ and $\beta_R$ are adjustable parameters. $\vec{M}$ is the mutual inductance matrix, which is measured:

$$M = \begin{pmatrix} 620 & -7 & -1.67 & 0 & \cdots & 0 & -1.67 & -7 \\ -7 & 620 & -7 & -1.67 & \cdots & 0 & 0 & -1.67 \\ -1.67 & -7 & 620 & -7 & \cdots & 0 & 0 & 0 \\ 0 & -1.67 & -7 & 620 & \cdots & 0 & 0 & 0 \\ \vdots & \vdots & \vdots & \vdots & \cdots & \vdots & \vdots & \vdots \\ 0 & 0 & 0 & 0 & \cdots & 620 & -7 & -1.67 \\ -1.67 & 0 & 0 & 0 & \cdots & -7 & 620 & -7 \\ -7 & -1.67 & 0 & 0 & \cdots & -1.67 & -7 & 620 \end{pmatrix} \mu H$$

The features above was the main reference of the feedback control system. These indispensable parameters were used as initialization parameters.

### B. Discrete PID controller

A PID (proportional–integral–derivative) controller is a loop feedback mechanism widely used in industrial control systems and a variety of other applications. This algorithm continuously calculated an error value e(t) as the difference between a set point and a measured process variable and applied a correction based on three different terms (denoted P, I, and D respectively) which gave the controller its name [6]. The control function can be expressed mathematically as

$$u(t) = K_p e(t) + K_i \int_0^t e(\tau)d\tau + K_d \frac{de(t)}{dt} \quad (2)$$

Kp is the proportional gain, Ki is the integral gain, Kd is the derivative gain.

Approximation with a sampling time △t for the integral part and derivative part can be replaced as

$$u(t_k) = u(t_{k-1}) + K_P \left[\left(1 + \frac{\Delta t}{T_i} + \frac{T_d}{\Delta t}\right)e(t_k) + \left(-1 - \frac{2T_d}{\Delta t}\right)e(t_{k-1}) + \frac{T_d}{\Delta t}e(t_{k-2})\right] \quad (3)$$

$T_I = K_p/K_I, T_d = K_d/K_p$

## IV. TEST RESULT

ALTERA Cyclone V is the core of both modules. ALTERA provides a debugging tool called SignalTap II that can be used to capture and display signals in real time in FPGA design. Using SignalTap II can show the status of component and solve problem easily. The Figure 6 shows the SignalTap ii waveform of the DAC, the ADC and the data transmission between the two boards. The sample clock for the SignalTap II is 20MHz.

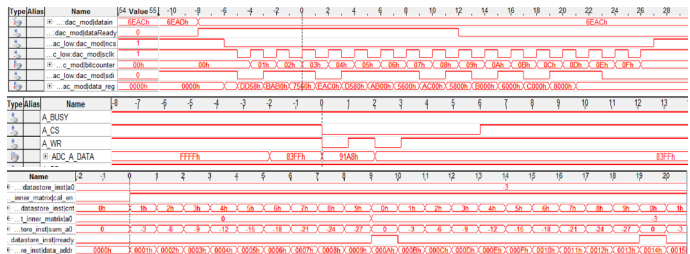

**Figure**.5 SignalTap II test result

From the figure, it can be concluded that both boards can work normally. The system can sample the analog signal and transmit a proper feedback signal to the power amplifier to reduce the error field of vertical gap. The figure below is the photograph of both boards, and the still some need to be done soon.

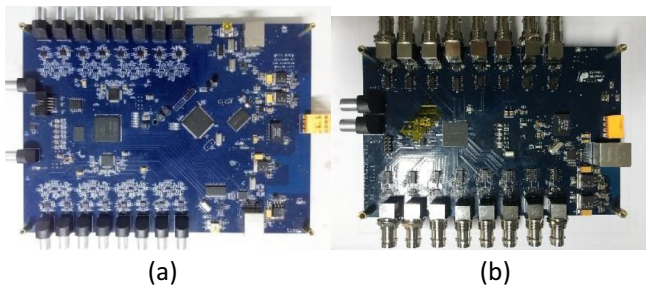

(a)             (b)

Figure.6. the photograph of both boards
(a)Sample module board (b)coil control module board

## V. CONCLUSION

This work focus on designing a feedback system to reduce the error field of vertical gap. FPGA is the core of the whole system. FPGA not only gathers and transmits data from rogovski coils, but also fulfill the calculation of two different algorithms in real time. The division of work is clear for both modules. Results indicated that this system could successfully realize mutual inductance correction and PID controller in real time.